\pdfoutput=1
\expandafter\def\csname ver@natbib.sty\endcsname{2099/01/01}
\def\bibpunct#1#2#3#4#5#6{}
\documentclass[pmlr,11pt]{jmlr}
\newenvironment{proofof}[1]{\begin{proof}\textbf{of {#1}}}{\end{proof}}
\expandafter\let\csname ver@natbib.sty\endcsname\relax
\makeatletter
\def\ps@jmlrtps{%
  \let\@mkboth\@gobbletwo
  \def\@oddhead{}\let\@evenhead\@oddhead
  \def\@oddfoot{\hfill\small\rmfamily\thepage\hfill}\let\@evenfoot\@oddfoot}
\makeatother
\makeatletter
\AtBeginDocument{\renewcommand*{\@jmlrenddoc}{%
  \FloatBarrier\phantomsection
  \protected@edef\@currentlabelname{end of \@shorttitle}%
  \label{jmlrend}\global\let\@reprint\@empty}}
\makeatother

\usepackage{times}
\usepackage{mathtools}
\usepackage{booktabs}
\usepackage{microtype}
\usepackage{csquotes}
\usepackage[
  backend=biber,
  style=authoryear-comp, sortcites=false,
  natbib=true,
  maxcitenames=2,
  maxbibnames=99,
  uniquename=false,
  uniquelist=false,
  giveninits=true,
  doi=true,
  url=false,
  isbn=false,
  eprint=true,
  date=year
]{biblatex}
\AtEveryBibitem{%
  \clearfield{urlyear}\clearfield{urlmonth}\clearfield{urlday}%
  \clearfield{version}\clearfield{pubstate}\clearfield{issn}%
  \clearfield{eventtitle}\clearlist{language}\clearfield{langid}%
}
\DeclareSourcemap{
  \maps[datatype=bibtex]{
    \map{
      \step[fieldsource=entrykey,
            match=\regexp{\AkempaCollapsingHierarchyCompressed2023\Z}, final]
      \step[typesource=article, typetarget=online]
      \step[fieldset=eprinttype, fieldvalue={arXiv}]
      \step[fieldset=eprint, fieldvalue={2308.03635}]
    }
  }
}
\DeclareMathOperator{\poly}{poly}
\DeclareMathOperator{\polylog}{polylog}
\newcommand{\rev}[1]{{#1}^{R}}
\newcommand{\BWT}{\ifmmode\operatorname{BWT}\else\textup{BWT}\fi}
\newcommand{\RLBWT}{\ifmmode\operatorname{RLBWT}\else\textup{RLBWT}\fi}
\newcommand{\SA}{\ifmmode\operatorname{SA}\else\textup{SA}\fi}
\newcommand{\occ}{\operatorname{occ}}
\newcommand{\rank}{\operatorname{rank}}
\newcommand{\insertop}{\operatorname{insert}}
\newcommand{\cand}{\operatorname{candidate}}
\newcommand{\sumop}{\operatorname{sum}}
\newcommand{\searchop}{\operatorname{search}}
\newcommand{\Trank}{T_{\mathrm{rank}}}
\newcommand{\Tcount}{T_{\mathrm{count}}}
\newcommand{\Tins}{T_{\mathrm{ins}}}
\newcommand{\Tcand}{T_{\mathrm{cand}}}

\SetKwInput{KwState}{State}
\SetKwInput{KwInit}{Initially}
\SetKwProg{Proc}{}{}{}
\SetFuncSty{textsc}
\LinesNumbered
\DontPrintSemicolon

\csdef{jmlr@thm@definition@body@font}{\normalfont}
\AtBeginEnvironment{definition}{\csdef{jmlr@thm@theorem@body@font}{\normalfont}}

\newtheorem*{openproblem}{Open Problem}

\title[Efficient Lempel-Ziv Prediction]{Efficient Lempel-Ziv Prediction with the Run-Length BWT}
\hypersetup{pdftitle={Efficient Lempel-Ziv Prediction with the Run-Length BWT}}
\hypersetup{pdfauthor={Ohad Avnery}}

\author{%
 \Name{Ohad Avnery} \Email{ohad.avnery@gmail.com}\\
 \addr Computational Rational Agents Laboratory (CORAL)\\
}

\begin{document}

\maketitle

\begin{abstract}%
  Lempel-Ziv Plurality (LZP) is an algorithm of Kosoy that predicts the next
  symbol of a string by using its Lempel-Ziv factorization, and makes at most
  $O(z\log n)$ mistakes on every string of length $n$ with $z$ LZ77 phrases.
  We give efficient online implementations of LZP and of a new variant of it.
  Our first implementation maintains a run-length encoded
  Burrows-Wheeler transform (\RLBWT{}) of the reversed history. It implements
  LZP, uses $O(z\log z\log^2 n)$ bits, and has worst-case update and prediction times
  $O(\log z+\log\log n)$ and $O(\sigma(\log z+\log\log n))$, respectively, where
  $\sigma$ is the size of the alphabet. This is an exponential
  improvement over Kosoy's implementation, whose update time is $O(z\log^7 n)$.
  We then introduce the Lempel-Ziv Majority (LZM) predictor, which has the same mistake bound and update time, and predicts in $O(1)$ time independently of $\sigma$. These two predictors accept new symbols at any time, without knowing the
  alphabet in advance. For an alphabet known in advance, a different data structure gives $O(\sigma(1+\log z/\log\log n))$
  time per symbol, which is constant when $\sigma=O(1)$ and
  $z=\polylog(n)$.%
\end{abstract}

\begin{keywords}%
  online learning, sequence prediction, mistake bounds, Lempel-Ziv
  factorization, Burrows-Wheeler transform, dynamic data structures%
\end{keywords}

\section{Introduction}
\label{sec:intro}

Online sequence prediction is a classical problem in machine learning and
algorithms. A predictor reads a sequence one symbol at a time, each time
guessing the next symbol, then reading the true symbol and updating its
state.
What one would like is a single algorithm that is at once computationally efficient,
general-purpose (able to learn a wide range of patterns from many domains),
and equipped with a formal mistake bound that guarantees a small number of
prediction mistakes for the sequences we care about. Unfortunately, we do
not know of any predictor that satisfies all three properties.

The theoretical gold standard for sequence prediction is Solomonoff
induction \citep{hutterIntroductionUniversalArtificial2024}, which gives higher weight to sequences with
small Kolmogorov complexity: those generated by short computer programs.
This predictor satisfies the second and third requirements, with
guaranteed mistake bounds on all computable data; however, Solomonoff
induction is itself uncomputable, as is the notion of Kolmogorov complexity
it is based on.
In order to build efficient predictors, we should relax the notion of string
complexity to simpler patterns that are easier to recognize efficiently.

A natural repetitiveness measure is the size of the LZ77 parse: for a string
$x$, $z(x)$ is defined to be the number of phrases in its LZ77 factorization.
This measure captures highly repetitive strings, which can be built by
reusing and copying previous terms.
The broader Lempel-Ziv approach was in fact first proposed as a
computationally tractable stand-in for Kolmogorov complexity
\citep{lempelComplexityFiniteSequences1976a}, and it has considerable
expressive power: many rich families of deterministic sequences studied in
combinatorics on words \citep{alloucheAutomaticSequencesTheory2003}, such as
automatic sequences, Sturmian words and many morphic sequences, have LZ77
parse size only logarithmic in their length
\citep{kosoyStringologicalSequencePrediction2026,navarroIndexingHighlyRepetitive2022},
and are therefore well-predicted by any algorithm whose error scales with $z$.
Predicting strings with small $z(x)$ well can thus serve as a computable toy
analogue of the Solomonoff ideal.

Kosoy's Lempel-Ziv Plurality predictor (LZP) does exactly this
\citep{kosoyStringologicalSequencePrediction2026}. At each step it takes the
current LZ77 phrase, finds the earlier places in the history where that
phrase has occurred, and predicts the symbol that most often followed them.
LZP makes at most $O(z\log n)$ mistakes on any length-$n$ sequence, so a
highly repetitive sequence is predicted with correspondingly few errors.
However, the given implementation of LZP reconstructs an auxiliary compressed
index on each symbol's update and prediction, giving $O(z\log^7 n)$ time for each
operation when the alphabet is constant
\citep[Section~D.2.3]{kosoyStringologicalSequencePrediction2026}.

Our first contribution removes this cost. Rather than storing the LZ77
factorization explicitly, we maintain a run-length encoded Burrows-Wheeler
transform (\RLBWT{}) of the reversed history. In this representation, all
occurrences of a given phrase form a single contiguous interval, and the
symbols that follow them can be counted efficiently using \emph{rank} queries. Extending the current phrase by
one symbol is a single incremental update. This implements LZP, while reducing the time of a single update to
$O(\log z+\log\log n)$.

Our second contribution changes the prediction rule itself to make prediction faster. LZP prediction calculates the exact plurality in a range by iterating over all distinct symbols, costing an extra factor of the alphabet size $\sigma$. We define a weaker prediction rule that is only guaranteed to predict a symbol if it has over half of the votes. The resulting Lempel-Ziv Majority predictor (LZM) inherits the same mistake bound as LZP. It calculates the vote majority with a single query, using Boyer-Moore cancellation. This allows prediction to run as fast as updates, and precalculating prediction makes prediction time $O(1)$.

Our third contribution gives a faster implementation for small, fixed alphabets. Our predictors only need a small interface of a dynamic run-length encoded string that the dynamic RLBWT is implemented over. Each RLBWT update only requires local changes to the runs. This allows us to use the dynamic partial sums structure of \citet{billeDynamicRelativeCompression2018}. The resulting LZP implementation does updates and predictions in $O(\sigma(1+\log z/\log\log n))$ worst-case time. In particular, we achieve worst-case \emph{constant} time when $\sigma=O(1)$ and $z=\polylog(n)$.

None of these implementations is fastest for all parameters
(\tableref{tab:comparison}). In \sectionref{sec:open} we ask for a single
dynamic run-length encoded string with range-majority queries whose
operations all take $O(1+\log r/\log w)$ worst-case time, where $r$ is the
number of runs and $w$ is the word size. Such a
structure would make LZM run in constant time per symbol whenever
$z=\polylog(n)$, for any alphabet. We believe that such a structure exists, and leave a construction for future work.

\begin{table}[t]
  \centering
  \caption{Lempel-Ziv predictors for $w=\Theta(\log n)$, with $\sigma$
  as defined in \sectionref{sec:setting}. All
  make at most $z(\log_2 n+1)$ mistakes. Our rows are
  \corollaryref{cor:main}\,(1)--(3): they are deterministic, with worst-case
  times and space including temporary storage.
  First row follows \citet[Sections~D.2.1 and~D.2.3]{kosoyStringologicalSequencePrediction2026},
  assuming $\sigma=O(1)$ and counting only persistent state.}
  \label{tab:comparison}
  \footnotesize
  \setlength{\tabcolsep}{3pt}
  \begin{tabular}{@{}lllll@{}}
    \toprule
    Predictor & Update & Prediction & Space (bits) & Alphabet \\
    \midrule
    LZP (Kosoy)
      & $O(z\log^7 n)$
      & $O(z\log^7 n)$
      & $O(z\log n)$
      & fixed \\
    \RLBWT{}-LZP
      & $O(\log z+\log\log n)$
      & $O(\sigma(\log z+\log\log n))$
      & $O(z\log z\log^2 n)$
      & dynamic \\
    \RLBWT{}-LZM
      & $O(\log z+\log\log n)$
      & $O(1)$
      & $O(z\log z\log^2 n)$
      & dynamic \\
    \RLBWT{}-LZP, alphabet-indexed
      & $O(\sigma(1+\log z/\log\log n))$
      & $O(1)$
      & $O(\sigma z\log z\log^2 n)$
      & fixed \\
    \bottomrule
  \end{tabular}
  \par\smallskip
\end{table}

\subsection{Related Work}
\label{sec:related}

Universal prediction of individual sequences has long been connected to Lempel-Ziv parsing.
\citet{federUniversalPredictionIndividual1992} give a randomized LZ78-based predictor and bound its asymptotic expected fraction of mistakes on every infinite binary sequence. We instead consider finite mistake bounds in terms of the number of LZ77 phrases.
\citet{kosoyStringologicalSequencePrediction2026} introduced LZP and proved such a bound, within a framework of stringological sequence prediction that is developed further in \citet{kosoyStringologicalSequencePrediction2026a,kosoyStringologicalSequencePrediction2026iii}.

Compressed string indexes have also been used for prediction: \citet{ktistakisSuccinctBWTBasedSequence2019} use the FM-index and
\citet{cunialFrameworkSpaceefficientVariableorder2019} use a run-length \BWT{}, both for predicting the next symbol of a sequence. Both approaches build indexes on fixed training data and neither establishes mistake bounds. Our predictors instead maintain a dynamic index of the growing history and satisfy finite mistake bounds in terms of the number of LZ77 phrases.

\citet{policritiComputingLZ77RunCompressed2016} and \citet{ohnoFasterImplementationOnline2018} use dynamic run-length \BWT{}s to compute the LZ77 factorization in compressed space, \citet{bannaiOnlineLZ77Parsing2018} use them for online LZ77 parsing, and \citet{nishimotoOptimalTimeRLBWTConstruction2022} construct the \RLBWT{} online in $O(n+r\log r)$ total time. We follow the layout of \citet{policritiComputingLZ77RunCompressed2016,ohnoFasterImplementationOnline2018}, implemented with the dynamic sequences of \citet{navarroOptimalDynamicSequence2014} and the dynamic partial sums of \citet{billeDynamicRelativeCompression2018}. Our predictors use \BWT{} \emph{rank} queries to obtain continuation counts, and LZM uses an additional range-majority query, all with worst-case bounds per symbol. The bound of \citet{kempaResolutionBurrowsWheelerTransform2020} on the number of \BWT{} runs relates our costs to $z$.

For LZM, we use a weighted form of the Boyer-Moore majority vote \citep{boyerMJRTYFastMajority1991}.
Dynamic range-majority queries have also been studied for general sequences \citep{elmasryDynamicRangeMajority2012},
including entropy-compressed ones \citep{gagieCompressedDynamicRange2020}; our structure works on run-length encoded sequences, and weights runs by their lengths.

\section{Setting}
\label{sec:setting}

We recall the prediction framework of
\citet{kosoyStringologicalSequencePrediction2026}, fix our model of
computation, and define the two prediction rules that we implement, which
share the mistake bound of LZP.
Proofs omitted from this section are in \appendixref{app:setting}.

\subsection{Notation and Model of Computation}
\label{sec:notation}

Strings are indexed from zero: for a string $u$ we write $u[i]$ for its $i$-th
symbol, $u[i:j]\coloneqq u[i]{u[i+1]}\cdots {u[j-1]}$, $u[:j]\coloneqq u[0:j]$ and $u[i:]\coloneqq u[i:|u|]$.
The reverse of $u$ is $\rev{u}\coloneqq {u[|u|-1]}\cdots u[1]u[0]$. For strings
$W$ and $H$, $\occ_H(W)$ is the number of occurrences of $W$ in $H$, i.e.\ the
number of positions $j$ with $H[j:j+|W|]=W$. For a condition $\varphi$,
$[\varphi]$ is $1$ if $\varphi$ holds and $0$ otherwise. Logarithms are to
base $2$, and inside $O$-notation we let $\log x\coloneqq\max\{1,\log_2 x\}$
for $x>0$ and $\log 0\coloneqq 1$.

We use the word RAM with a word size $w$ that is fixed throughout an
execution, and with the usual constant-time arithmetic (including
multiplication), comparison, Boolean and shift operations. For an input of
length $n$, we assume that $w=\Theta(\log n)$, chosen large enough that
positions, counts and pointers fit in a word. The
space of an algorithm is the largest number of words that it occupies at any
time, including temporary storage. We state space bounds in words, and
translate them to bits for $w=\Theta(\log n)$. All algorithms in this paper
are deterministic. All time bounds are worst-case rather than amortized.

The alphabet $\Sigma$ can be either fixed or dynamic. In the fixed settings, $\sigma = |\Sigma|$ is known ahead of time and the time and space bounds allow dependency on $\sigma$. We set $\Sigma=\{0,1,\dots,\sigma-1\}$, and we assume $\sigma \le 2^w$ so that each symbol fits in a single word. In the dynamic settings, all $w$-bit words are valid symbols, $\Sigma = \{0,1,\dots,2^w - 1\}$.
Here $\sigma$ denotes the number of distinct symbols observed so far;
for bounds on a complete input $x$, it denotes the number of distinct symbols in $x$.

\subsection{The Prediction Protocol}
\label{sec:protocol}

Following \citet{kosoyStringologicalSequencePrediction2026}, a
\emph{predictor} is a tuple $\Pi=(\mathcal S,s_{\mathrm{init}},\mathcal
U,\mathcal P)$, where $\mathcal S\subseteq\{0,1\}^*$ is the set of internal
states, $s_{\mathrm{init}}\in\mathcal S$ is the initial state, $\mathcal
U:\mathcal S\times\Sigma\to\mathcal S$ is the state-update function and
$\mathcal P:\mathcal S\to\Sigma$ is the state-prediction function. On an input
$x\in\Sigma^n$, the predictor starts at $s_0\coloneqq s_{\mathrm{init}}$ and,
at each round $t=0,1,\dots,n-1$, predicts $\hat x_t\coloneqq\mathcal P(s_t)$,
observes $x[t]$, and updates its state to $s_{t+1}\coloneqq\mathcal
U(s_t,x[t])$. The number of mistakes and the maximal state size are
\[
  M_\Pi(x)\coloneqq\bigl|\{t<n \mid \hat x_t\ne x[t]\}\bigr|,
  \qquad
  S_\Pi(x)\coloneqq\max_{0\le t<n}|s_t|.
\]
\citet{kosoyStringologicalSequencePrediction2026} calls a predictor
\emph{statistically efficient} with respect to a complexity measure $C$ if
$M_\Pi(x)\le(C(x)+1)\poly(\log|x|)$ for all $x$, and \emph{computationally
efficient} if each round takes time polynomial in the size of the state and
$S_\Pi(x)\le\poly(C(x),\log|x|)$. For every fixed alphabet, our predictors are statistically and computationally efficient with respect to $z$.

The \emph{update time} and \emph{prediction time} of a predictor are the
worst-case times of computing $\mathcal U$ and $\mathcal P$, respectively.
Any predictor can be \emph{cached}: the update computes
$\mathcal P(s_{t+1})$ right after $s_{t+1}$ and stores it in the state, so
that a prediction only reads the stored symbol. This adds the prediction time
to the update time, makes the prediction time $O(1)$, and adds one word to
the state; the predictions, and hence the mistakes, are unchanged.

\subsection{Lempel-Ziv Factorization and LZP}
\label{sec:lzp}

\begin{definition}[LZ77 factorization]
\label{def:lz77}
The LZ77 factorization of a string $u\in \Sigma^*$ is the unique decomposition $u=f_0f_1\cdots f_{z-1}$ into phrases, such that each $f_i$ is either a symbol $c\in \Sigma$ that does not occur earlier in $u$, or the longest prefix of $u[|f_0\cdots f_{i-1}|:]$ that has an earlier occurrence in $u$.
The LZ77 complexity $z(u)$ is the number of phrases.
\end{definition}

A phrase $f_i$ may overlap its earlier occurrence: if $f_i$ starts at position $\ell$, it may be copied from a source
starting at any $j<\ell$, even if $j+|f_i|>\ell$. We denote $z:=z(x)$ for the input string $x$. We let the last phrase of the
empty string be the empty string. As $x$ is given online, its last phrase evolves as
follows.

\begin{lemma}[Phrase extension]
\label{lem:extension}
Let $H\in\Sigma^*$ have last phrase $P$, and let $c\in\Sigma$. If
$\occ_H(Pc)>0$, the factorization of $Hc$ is obtained from that of $H$ by
extending its last phrase to $Pc$. Otherwise, it is obtained by appending
the new phrase $c$.
\end{lemma}

For a history $H$ with last phrase $P$, the \emph{votes} for the next symbol are the
continuations of the earlier occurrences of $P$:
\[
  v_H(P,c)\coloneqq\occ_H(Pc)\quad\text{for } c\in\Sigma,
  \qquad
  v_H(P)\coloneqq\sum_{c\in\Sigma}v_H(P,c).
\]

\begin{definition}[{LZP; \citealp[Alg.~7]{kosoyStringologicalSequencePrediction2026}}]
\label{def:lzp}
On a history $H$ with last phrase $P$, LZP predicts a symbol
$c\in\Sigma$ maximizing $v_H(P,c)$, breaking ties arbitrarily.
\end{definition}

\begin{definition}[Majority-respecting predictors]
\label{def:lzm}
A predictor is \emph{majority-respecting} if, on every history $H$ with last
phrase $P$, it predicts $c$ whenever $v_H(P,c)>\tfrac{1}{2}v_H(P)$.
\end{definition}

LZP is majority-respecting, and so is the \emph{Lempel-Ziv Majority} (LZM)
predictor defined in \sectionref{sec:lzm} (\theoremref{thm:lzm}).

\begin{theorem}[Mistake bound]
\label{thm:mistakes}
Every majority-respecting predictor makes at most $z(\log_2 n+1)$ mistakes
on every $x\in\Sigma^n$ with $n\ge1$, where $z=z(x)$. In particular, this
holds for LZP and LZM.
\end{theorem}

The argument of
\citet[Appendix~D]{kosoyStringologicalSequencePrediction2026} for LZP uses
only the strict-majority property, so it also establishes the bound for
every majority-respecting predictor; for completeness, we include the full
argument in \appendixref{app:setting}.

\subsection{The Burrows-Wheeler Transform}
\label{sec:bwt}
Let $\$ \notin \Sigma$ be the \emph{sentinel} symbol and denote $\Sigma'=\Sigma \cup \{\$\}$. We order $\Sigma$ by numerical order and let $\$ < c$ for all $c\in \Sigma$, thus inducing a lexicographic ordering on $(\Sigma')^*$.

Let $Y=U\$$ be a sentinel-terminated string, where $U\in\Sigma^*$, and let $N\coloneqq|Y|$.

For such a string $Y$, its suffix array $\SA\in\{0,1,\dots,N-1\}^N$ is the permutation
of starting positions ordered by lexicographic ordering of the relevant
suffixes:
\[
  Y[\SA[0]:N]<Y[\SA[1]:N]<\dots<Y[\SA[N-1]:N].
\]

The Burrows-Wheeler transform \citep{burrowsBlocksortingLosslessData1994}
$L=\BWT(Y)$ is the string of symbols preceding the suffixes of $Y$ in
suffix-array order: $L$ has length $N$ and $L[j]=Y[\SA[j]-1]$, with wraparound
$Y[-1]\coloneqq Y[N-1]=\$$
\citep{ferraginaCompressedRepresentationsSequences2007,navarroIndexingHighlyRepetitive2022}.
We call each $j\in[0,N)$ a \emph{row}: row $j$ represents the suffix
$Y[\SA[j]:N]$, and its \BWT{} symbol is $L[j]$.\footnote{The name comes from the \emph{Burrows-Wheeler matrix},
whose rows are the cyclic rotations of $Y$ in lexicographic order: row $j$
starts with the suffix $Y[\SA[j]:N]$ and ends with $L[j]$. We do not use the
matrix itself.}

For $c\in\Sigma'$ and $0\le i\le N$, let $\rank_c(L,i)$ be the number of
occurrences of $c$ in $L$ up to (but not including) index $i$:
\[
  \rank_c(L,i)\coloneqq\bigl|\{q<i\mid L[q]=c\}\bigr|.
\]
Let $C_L(c)$ count the symbols of $L$ smaller than $c$:
\[
  C_L(c)\coloneqq\sum_{\substack{d\in\Sigma'\\d<c}}\rank_d(L,|L|)
    =\bigl|\{q<N\mid L[q]<c\}\bigr|.
\]

The set of all suffixes beginning with a given pattern $W\in\Sigma^*$
occupies a suffix-array interval $I_Y(W)=[a,b)$. For nonempty $W$, the
pattern has exactly $b-a$ occurrences in $Y$, and the interval is empty
when no such occurrences exist. For the empty pattern, we set
$I_Y(\varepsilon)=[0,N)$, containing all $N$ suffix-array rows.

\BWT{} supports efficient pattern matching in $Y$, by \emph{backward
search} \citep{ferraginaOpportunisticDataStructures2000,ferraginaIndexingCompressedText2005},
whose steps extend the pattern one symbol at a time: for
$c\in\Sigma$,
\begin{equation}
  \label{eq:backward-step}
  I_Y(cW)=\bigl[C_L(c)+\rank_c(L,a),\;C_L(c)+\rank_c(L,b)\bigr).
\end{equation}
A \emph{run} of $L$ is a maximal block of equal symbols, and $r(L)$ is the
number of runs. The run-length encoded \BWT{} (\RLBWT{}) stores $L$ as the sequence of its runs.
$r(L)$ is small for repetitive texts: it can be bounded in terms of the LZ77 complexity of the underlying text (see \lemmaref{lem:runs}).

\section{Dynamic Run-Length Encoded Strings}
\label{sec:rle}

A run-length encoding (RLE) represents a string
$L=c_1^{e_1}c_2^{e_2}\cdots c_r^{e_r}$, where $c_j\ne c_{j+1}$ and $e_j\ge1$,
by its \emph{run heads} $c_j$ and \emph{run lengths} $e_j$, using space
proportional to the number $r$ of runs rather than to the length of $L$. Our predictors will store their \BWT{} this way (\sectionref{sec:predictors}).
Everything LZP needs from the representation is captured by the following
interface, which we describe independently of its application. Proofs omitted from this section are in \appendixref{app:rle}.

\begin{definition}[Incremental RLE]
\label{def:rle}
An \emph{incremental RLE structure} maintains a string $L\in\Sigma^*$ of length $N$,
initially empty, supporting the following operations:
\begin{itemize}
  \item $\rank_c(L,i)=\bigl|\{q<i\mid L[q]=c\}\bigr|$, for $c\in\Sigma$ and $0\le i\le N$;
  \item $C_L(c)=\bigl|\{q<N\mid L[q]<c\}\bigr|$, for $c\in\Sigma$;
  \item $\insertop(i,c)$, for $0\le i\le N$ and $c\in\Sigma$, which replaces
    $L$ by $L[:i]\,c\,L[i:]$.
\end{itemize}
The queried symbols need not occur in $L$. We write $\Trank$, $\Tcount$ and
$\Tins$ for the worst-case times of the three operations, and $S$ for the
number of words used, including temporary storage. These are functions of
the current number of runs $r$ (and of $w$ and $\sigma$), nondecreasing in $r$; when $L$ is empty, we evaluate them at $r=1$.
\end{definition}

This interface is weaker than that of general dynamic RLE structures, which
also support \emph{access}, \emph{select} and deletions; our results only
rely on the operations above, so any structure implementing them suffices.

LZM (\sectionref{sec:lzm}) needs one additional query, which returns the majority symbol of a range
of $L$ whenever one exists.

\begin{definition}[Augmented incremental RLE]
\label{def:rle-aug}
An \emph{augmented incremental RLE structure} is an incremental RLE structure
that also supports the query
$\cand(a,b)\in\Sigma\cup\{\bot\}$, for $0\le a\le b\le N$, in worst-case time
$\Tcand$, with the following one-sided guarantee: if some $c\in\Sigma$ occurs
in $L[a:b]$ more than $(b-a)/2$ times, then $\cand(a,b)=c$; otherwise, the
answer may be arbitrary.
\end{definition}

The following lemma describes how an insertion changes the runs of $L$.

\begin{lemma}[Run edits]
\label{lem:run-edits}
An operation $\insertop(i,c)$ changes the runs of $L$ in one of three ways:
it extends a run of $c$ by one symbol; it splits a run of another symbol
$c'$ into two runs of $c'$, with a new run $c^1$ between them; or it inserts
a new run $c^1$ between two runs of symbols other than $c$, or at either end of
$L$. In particular, no run is ever deleted, and no two runs are merged.
\end{lemma}

Our implementations are built from the following two primitives.

\begin{definition}[Partial sums]
\label{def:partial-sums}
A \emph{partial-sums structure} maintains a sequence $v_1,\dots,v_s$ of
nonnegative integers under updates and insertions of entries, supporting the
queries $\sumop(k)\coloneqq v_1+\dots+v_k$ and $\searchop(t)$, the smallest $k$
with $\sumop(k)\ge t$.
\end{definition}

\begin{definition}[Dynamic sequence]
\label{def:dyn-seq}
A \emph{dynamic sequence} maintains a string $X\in\Sigma^*$ under the
operations $\rank_c(X,j)$ and $\insertop(j,c)$, defined as for $L$ in
\definitionref{def:rle}.
\end{definition}

\subsection{Balanced-Tree Implementation}
\label{sec:rle-trees}

We first implement an incremental RLE structure following the layout of
\citet{policritiComputingLZ77RunCompressed2016,ohnoFasterImplementationOnline2018},
with components that have worst-case time bounds.
We call the symbols that occur in $L$ \emph{active}.

The structure represents the string $L=c_1^{e_1}\cdots c_r^{e_r}$ by the following substructures:
\begin{itemize}
  \item $A$: a partial-sums structure over the run lengths
    $e_1,\dots,e_r$;
  \item $A_c$, for every active symbol $c$: a partial-sums structure over
    the lengths of the runs of $c$, in order;
  \item $L'\coloneqq c_1c_2\cdots c_r$: the string of run heads, stored as a
    dynamic sequence;\footnote{As a
    string, $L'$ is indexed from zero, so $L'[j-1]=c_j$, while runs and the
    entries of partial-sums structures are numbered from one.}
  \item $F$: a dictionary of the active symbols, ordered by $\Sigma$, with
    the number of occurrences of each symbol in $L$.
\end{itemize}

If $i=0$ or $c$ is not active, which we test using $F$, then $\rank_c(L,i)=0$.
Otherwise, to answer $\rank_c(L,i)$, let
$k\coloneqq A.\searchop(i)-1$ be the number of runs before
the run containing $L[i-1]$, whose symbol is $L'[k]$. Then
$A.\sumop(k)$ symbols precede it, and $\rank_c(L',k)$ of the
runs before it are runs of $c$, so
\begin{equation}
  \label{eq:rle-rank}
  \rank_c(L,i)=A_c.\sumop\bigl(\rank_c(L',k)\bigr)
    +\bigl(i-A.\sumop(k)\bigr)\,\bigl[L'[k]=c\bigr].
\end{equation}
Similarly, $C_L(c)$ is the sum of the occurrences stored in $F$ for the
symbols smaller than $c$.

\begin{lemma}[Balanced trees]
\label{lem:rle-trees}
There is an incremental RLE structure with $\Trank,\Tcount,\Tins=O(\log r)$ and
$S=O(r)$ words, which supports symbols that are not known in advance.
\end{lemma}

The partial-sums structures and $F$ are balanced search trees with subtree
sums. For $L'$, which needs \emph{rank} queries over a growing alphabet, we
use the worst-case dynamic sequences of
\citet[Theorem~10.5]{navarroOptimalDynamicSequence2014}.

\subsection{Alphabet-Indexed Implementation}
\label{sec:rle-indexed}

We now give a second incremental RLE structure, for an alphabet
$\Sigma=\{0,\dots,\sigma-1\}$ known in advance. It is faster than the structure of \lemmaref{lem:rle-trees} when
$\sigma$ is small, e.g.\ $\sigma=O(1)$, but its insertion time and its space
grow linearly with $\sigma$.

By \lemmaref{lem:run-edits}, an insertion changes the sequence of run lengths
only by adding one to an entry, inserting an entry equal to one, or dividing
an entry into two entries with the same total. Such changes can be
implemented faster than with balanced trees, using the following result of
\citeauthor{billeDynamicRelativeCompression2018}.

\begin{lemma}[Partial sums; adapted from \citealp{billeDynamicRelativeCompression2018}]
\label{lem:dps}
There is a partial-sums structure that maintains a sequence of $s$
nonnegative $w$-bit integers in $O(s+1)$ words, and supports $\sumop$ and
$\searchop$ queries, adding one to an entry, inserting an entry equal to $0$
or $1$, and dividing an entry into two consecutive entries with the same
total, each in $O(1+\log s/\log w)$ worst-case time.
\end{lemma}

\begin{samepage}
The partial-sums structures of this lemma do not store the run heads, so we
use one of them per symbol. We keep:
\begin{itemize}
  \item $A$: a partial-sums structure of \lemmaref{lem:dps} over the run
    lengths $e_1,\dots,e_r$;
  \item $W_c$, for every $c\in\Sigma$: a partial-sums structure of the same
    kind and length as $A$, whose $j$-th entry is $e_j$ if $c_j=c$, and $0$
    otherwise;
  \item $\mathrm{cnt}$: an array whose entry $\mathrm{cnt}[c]$ is $C_L(c)$,
    for every $c\in\Sigma$.
\end{itemize}
\end{samepage}
We have $\rank_c(L,0)=0$. For $i>0$, let $k\coloneqq A.\searchop(i)-1$ as
before. Then, for every $c\in\Sigma$, including symbols that do not occur
in $L$,
\[
  \rank_c(L,i)=W_c.\sumop(k)+\bigl(i-A.\sumop(k)\bigr)\,
  \bigl[W_c.\sumop(k+1)>W_c.\sumop(k)\bigr].
\]
Hence a \emph{rank} query uses four operations of \lemmaref{lem:dps}, on
$A$ and $W_c$ only, and takes $O(1+\log r/\log w)$ time, independently of
$\sigma$.
An insertion finds the symbols of the runs next to position $i$ by probing
all the structures $W_c$, and repeats each of its at most two run edits in all of
them, with zero entries in the structures of the other symbols. A query
$C_L(c)$ reads one entry of $\mathrm{cnt}$, and inserting $c$ adds one to
the entries $\mathrm{cnt}[y]$ with $y>c$, in $O(\sigma)$ time.

\begin{lemma}[Alphabet-indexed structure]
\label{lem:rle-indexed}
For an alphabet $\Sigma=\{0,\dots,\sigma-1\}$ that is known in advance, there
is an incremental RLE structure with $\Trank=O(1+\log r/\log w)$,
$\Tcount=O(1)$, $\Tins=O(\sigma(1+\log r/\log w))$, and
$S=O(\sigma(r+1))$ words.
\end{lemma}

\subsection{Range Majority with Cancellation Summaries}
\label{sec:rle-majority}

To answer \emph{candidate} queries, we augment the balanced-tree implementation of \sectionref{sec:rle-trees} with Boyer-Moore summaries
\citep{boyerMJRTYFastMajority1991}.

Boyer and Moore explain their majority algorithm as a floor fight: each
voter knocks down one voter of a different candidate and is knocked down in
turn. When no further pairs remain, all voters left standing support the
same candidate, and a candidate with a strict majority always has voters
left. Their algorithm runs this fight left to right, keeping only the
surviving candidate and the number of survivors.

We apply this to the tree $A$, which has one node per run. A run $c^e$ has no inner cancellation and contributes $e$ votes for $c$ at once. Fights on disjoint parts of $L$ can be held separately and their outcomes merged: if $p$ remaining votes for $c$ meet $q$ remaining votes for $c'\neq c$, another $\min\{p,q\}$ pairs cancel out. So each node stores the outcome of the fight within its subtree, and a query merges $O(\log r)$ such outcomes.

Formally, a \emph{summary} is either $\bot$ or a
pair $(c,p)$ with $c\in\Sigma$ and $p\ge1$. It is \emph{valid} for a
multiset $M$ of symbols if $M$ can be partitioned into pairs of distinct
symbols together with $p$ copies of $c$ (or with nothing, for $\bot$). The
summary of a (possibly partial) run $c^e$ is $(c,e)$. Two
summaries are combined by
\begin{equation}
  \label{eq:combine}
  (c,p)\star(c',q)\coloneqq
  \begin{cases}
    (c,p+q)  & \text{if } c=c',\\
    (c,p-q)  & \text{if } c\ne c' \text{ and } p>q,\\
    (c',q-p) & \text{if } c\ne c' \text{ and } q>p,\\
    \bot     & \text{if } c\ne c' \text{ and } p=q,
  \end{cases}
\end{equation}
and $\bot$ is neutral: $\bot\star s=s\star\bot=s$.

\begin{lemma}[Cancellation summaries]
\label{lem:summaries}
If $s_1$ and $s_2$ are valid summaries for $M_1$ and $M_2$, then $s_1\star s_2$ is a
valid summary for $M_1\uplus M_2$. If a symbol $c$ occurs in $M$ more than $|M|/2$
times, then every valid summary of $M$ is $(c,p)$ for some $p\ge1$.
\end{lemma}

\begin{proof}
If the two candidates are equal, their uncancelled copies add up. Otherwise,
we pair $\min\{p,q\}$ copies of the first candidate with copies of the
second, and the rest of the larger side remains uncancelled. For the second
claim, let $c$ occur $m_c>|M|/2$ times. A pair of distinct symbols contains at
most one copy of $c$ and at least one of the $|M|-m_c$ other symbols, so at
least $m_c-(|M|-m_c)>0$ copies of $c$ are left uncancelled. Since all
uncancelled symbols are equal, the summary is $(c,p)$ with $p\ge1$.
\end{proof}

The operation $\star$ is not associative: for the run sequence $a^3 d^2 b^3 d^2 c^3$, the final summary when combining left-to-right is $(c,3)$, while combining right-to-left gives $(a,3)$. The same example shows that the summary does not always compute the plurality, as $d$ is the unique most common symbol but no parenthesization leads to $d$ as the winning summary.

Each node of the tree stores the summary of its subtree, obtained by
combining the summaries of its left subtree, its own run and its right
subtree. Like a subtree sum, it depends only on the node and its children,
so an insertion recomputes it only at the nodes on the path from the changed
nodes to the root and at the nodes changed by rebalancing.

A query $\cand(a,b)$ is then computed like a range-sum query in a tree with
subtree sums, with $\star$ in place of $+$ (\algorithmref{alg:candidate} in
\appendixref{app:rle}): it combines the summaries of the two partial runs at
the ends of $L[a:b]$ and of $O(\log r)$ whole runs and subtrees between them,
and returns the candidate symbol of the result, or $\bot$ for the empty
summary. By \lemmaref{lem:summaries}, this
satisfies the guarantee of \definitionref{def:rle-aug}.

\begin{lemma}[Majority tree]
\label{lem:rle-majority}
The structure of \lemmaref{lem:rle-trees} can be extended to an augmented
incremental RLE structure with $\Tcand, \Trank,\Tcount,\Tins=O(\log r)$ using $S=O(r)$ words.
\end{lemma}

\section{Lempel-Ziv Prediction with a Dynamic RLBWT}
\label{sec:predictors}

We now implement LZP and LZM on top of the interfaces of \sectionref{sec:rle}.
Both predictors maintain the \BWT{} of the reversed history, together with
the suffix-array interval of the reversed current phrase.
Proofs omitted from this section are in \appendixref{app:predictors}.

\subsection{Maintaining the RLBWT of the Reversed History}
\label{sec:maintain}

At round $t$, let $H\coloneqq x[:t]$ be the history, $Y\coloneqq\rev{H}\$$,
and $L\coloneqq\BWT(Y)$, of length $N=t+1$. When a symbol $c$ is read, $Y$
becomes $cY=\rev{(Hc)}\$$. Reversing the history turns appending into prepending, which is the direction
in which the \BWT{} can be updated online
\citep{policritiComputingLZ77RunCompressed2016,ohnoFasterImplementationOnline2018}.
Prepending $c$ changes $L$ in two places: in the row of $Y$, the symbol $\$$
becomes $c$, and a new row for $cY$, with symbol $\$$, is inserted.
Since our interface supports neither deletions nor
replacements, we keep the sentinel outside of the structure. Let $p_\$$ be
the position of $\$$ in $L$, and let $L^-$ be the string over $\Sigma$
obtained from $L$ by deleting that position. The predictors store $L^-$ in an
incremental RLE structure, together with the number $p_\$$.

\begin{samepage}
\begin{lemma}[Sentinel-free maintenance]
\label{lem:sentinel}
Let $L$, $L^-$ and $p_\$$ be as above.
\begin{enumerate}
  \item For $c\in\Sigma$ and $0\le i\le N$,
    $\rank_c(L,i)=\rank_c(L^-,i-[i>p_\$])$ and $C_L(c)=C_{L^-}(c)+1$.
  \item Replacing $Y$ by $cY$, for $c\in\Sigma$, amounts to the operation
    $\insertop(p_\$,c)$ on $L^-$, followed by the assignment $p_\$\gets s$
    for
    \begin{equation}
      \label{eq:new-row}
      s\coloneqq C_{L^-}(c)+\rank_c(L^-,p_\$)+1,
    \end{equation}
    computed before the insertion; $s$ is the row of the new suffix $cY$.
\end{enumerate}
Hence, \emph{rank} and $C_L$ queries on $L$ take $O(\Trank)$ and $O(\Tcount)$
time, and prepending a symbol takes $O(\Trank+\Tcount+\Tins)$ time.
\end{lemma}
\end{samepage}

The costs of our structures depend on the number of runs of the maintained
\BWT{}. For $x\in\Sigma^n$, let
\[
  r_{\max}(x)\coloneqq\max_{0\le t\le n} r\bigl(\BWT(\rev{x[:t]}\$)\bigr).
\]
Deleting the sentinel does not increase the number of runs, so $L^-$ has at
most $r_{\max}(x)$ runs throughout the execution. We bound $r_{\max}(x)$ in
\lemmaref{lem:runs}.

\subsection{RLBWT-LZP}
\label{sec:lzp-impl}

\begin{algorithm2e}[htb]
\caption{\RLBWT{}-LZP}
\label{alg:lzp}
\KwState{the incremental RLE structure over $L^-$ and $p_\$$; the endpoints
  $a,b$ of the suffix-array interval $I=[a,b)\coloneqq I_Y(\rev{P})$, where $P$ is the last
  phrase of $H$; a linked list of
  the active symbols.}
\KwInit{$L^-$ is empty, $p_\$=0$, $I=[0,1)$, and the list of active symbols
  is empty. (This represents the empty history: $Y=\$$, $L=\$$ and
  $I=I_Y(\varepsilon)$.)}
\BlankLine
\Proc{\FuncSty{Update}$(c)$, when the symbol $c$ is read}{
  $J=[a',b')\gets I_Y(c\rev{P})$, by the backward step
    \eqref{eq:backward-step} from $I$\;
  \lIf{$\rank_c(L,N)=0$}{append $c$ to the list of active symbols\label{line:lzp-active}}
  update $L^-$ and $p_\$$ to prepend $c$ to $Y$, as in \lemmaref{lem:sentinel}\;
  \eIf{$J$ is nonempty}{
    $I\gets[a',b'+1)$\;\label{line:lzp-extend}
  }{
    $I\gets I_{cY}(c)$, by a backward step from the full interval $[0,N+1)$\;
  }
}
\BlankLine
\Proc{\FuncSty{Predict}}{
  \eIf{the list of active symbols is empty}{
    predict an arbitrary symbol of $\Sigma$\;
  }{
    $\nu_c\gets\rank_c(L,b)-\rank_c(L,a)$ for every active symbol $c$\;
    predict an active symbol $c$ with the largest $\nu_c$\;
  }
}
\end{algorithm2e}

\algorithmref{alg:lzp} gives \RLBWT{}-LZP.
The last step of the update follows \lemmaref{lem:extension}: $J$ is nonempty if and only if $Pc$ occurs in $H$, and then the phrase becomes $Pc$.
To see that the new interval is $[a',b'+1)$ in line~\ref{line:lzp-extend}, note that prepending $c$ to $Y$ inserts a single row, that of the new suffix $cY$, and keeps the relative order of the old suffixes.
Since $Y$ starts with $\rev{P}$, the new suffix $cY$ starts with
$c\rev{P}$, so its row $s$ satisfies
$a'\le s\le b'$. Hence the rows before $a'$ do not move, and the suffixes
starting with $c\rev{P}$ are the $b'-a'$ old ones and $cY$.

Each row of $I$ corresponds to an occurrence of $\rev{P}$ in $Y$, and hence
of $P$ in $H$. Its \BWT{} symbol precedes the occurrence in $Y$, so it is the
symbol \emph{following} the occurrence in $H$. The exception is the suffix $P$
of $H$, whose row is $p_\$$, with symbol $\$$. Counting the rows of $I$ by
their symbols therefore gives the votes of LZP:

\begin{lemma}[Votes]
\label{lem:votes}
Let $H$ have last phrase $P$, let $I=[a,b)\coloneqq I_Y(\rev{P})$, and for $c\in\Sigma$ let
$\nu_c\coloneqq\rank_c(L,b)-\rank_c(L,a)$. Then $\nu_c=v_H(P,c)$ for every $c\in\Sigma$. Moreover, $a\le p_\$<b$ and
$\sum_{c\in\Sigma}\nu_c=b-a-1$.
\end{lemma}

Since only active symbols can have positive votes, an active symbol with the
largest $\nu_c$ maximizes $v_H(P,c)$ over all of $\Sigma$, so the prediction
is that of LZP. (If all votes are zero, every symbol is tied.) Unlike online LZ77 parsing with an
\RLBWT{} \citep{bannaiOnlineLZ77Parsing2018}, the predictor never needs to
know \emph{where} the occurrences of $P$ are, and hence needs no suffix-array samples.

\begin{theorem}[\RLBWT{}-LZP]
\label{thm:lzp}
Instantiate \textup{\RLBWT{}-LZP} with an incremental RLE structure of
\definitionref{def:rle}. On every $x\in\Sigma^n$, it implements LZP, and
hence makes at most $z(\log_2 n+1)$ mistakes. Its update
time is $O(\Trank+\Tcount+\Tins)$ and its prediction time is
$O(\sigma\Trank)$; with caching,
the update time is $O(\sigma\Trank+\Tcount+\Tins)$ and the prediction time is
$O(1)$. Its space is $S+O(\sigma)$ words. The costs are evaluated at the
largest number of runs $r_{\max}(x)$.
\end{theorem}

\subsection{RLBWT-LZM}
\label{sec:lzm}

\begin{algorithm2e}[htb]
\caption{\RLBWT{}-LZM}
\label{alg:lzm}
\KwState{the \emph{augmented} incremental RLE structure over $L^-$ and
  $p_\$$; the endpoints of the interval $I=[a,b)$, as in \algorithmref{alg:lzp}.}
\KwInit{as in \algorithmref{alg:lzp}.}
\BlankLine
\Proc{\FuncSty{Update}$(c)$, when the symbol $c$ is read}{
  as in \algorithmref{alg:lzp}, except line~\ref{line:lzp-active}\;
}
\BlankLine
\Proc{\FuncSty{Predict}}{
  $\hat c\gets\cand(a,b-1)$ on $L^-$\;
  \eIf{$\hat c\ne\bot$}{
    predict $\hat c$\;
  }{
    predict an arbitrary symbol of $\Sigma$\;
  }
}
\end{algorithm2e}

\RLBWT{}-LZM (\algorithmref{alg:lzm}) maintains the same $L^-$, $p_\$$ and
$I$ as \RLBWT{}-LZP, but stores $L^-$ in an augmented incremental RLE
structure (\definitionref{def:rle-aug}), and it needs no list of active
symbols.
By \lemmaref{lem:votes}, the votes are the \BWT{}
symbols of the rows of $I=[a,b)$ other than the sentinel row $p_\$$. In
$L^-$, where the sentinel row is deleted, they occupy the range $[a,b-1)$.
If some symbol has a strict majority of the votes, then it has a strict
majority in $L^-[a:b-1]$, and the candidate query returns it. The sentinel
has to stay out of the query: for a history $H=cc$, the last phrase is $c$
with the single vote $c$, and the rows of $I$ have the \BWT{} symbols $c\$$,
so a summary that treated $\$$ as a symbol would cancel the only vote.

\begin{theorem}[\RLBWT{}-LZM]
\label{thm:lzm}
Instantiate \textup{\RLBWT{}-LZM} with an augmented incremental RLE structure of
\definitionref{def:rle-aug}. It is majority-respecting, and hence makes at most
$z(\log_2 n+1)$ mistakes on every $x\in\Sigma^n$. Its update time is
$O(\Trank+\Tcount+\Tins)$ and its prediction time is $O(\Tcand)$; with
caching, the update time is $O(\Trank+\Tcount+\Tins+\Tcand)$ and the
prediction time is $O(1)$. Its space is $S+O(1)$ words. The costs are
evaluated at $r_{\max}(x)$.
\end{theorem}

\subsection{Main Results}
\label{sec:main}

\theoremref{thm:lzp} and \theoremref{thm:lzm} bound the costs in terms of
$r_{\max}(x)$. To state our results in terms of $z$, we use the following
bound.

\begin{lemma}[Runs; based on \citealp{kempaResolutionBurrowsWheelerTransform2020}]
\label{lem:runs}
For every $x\in\Sigma^n$ with $n\ge1$, $r_{\max}(x)=O(z\log z\log n)$, where
$z=z(x)$.
\end{lemma}

Combining the black-box theorems with the implementations of
\sectionref{sec:rle} and \lemmaref{lem:runs} gives our main results.

\begin{corollary}[Main results]
\label{cor:main}
Let $x\in\Sigma^n$ with $n\ge2$, let $z=z(x)$, and let $w=\Theta(\log n)$. All of the following
predictors make at most $z(\log_2 n+1)$ mistakes on $x$.
\begin{enumerate}
  \item \textup{\RLBWT{}-LZP} with the balanced-tree structure of
    \lemmaref{lem:rle-trees} implements LZP. Its update
    time is $O(\log z+\log\log n)$, its prediction time is
    $O(\sigma(\log z+\log\log n))$, and its space is $O(z\log z\log n)$ words,
    i.e.\ $O(z\log z\log^2 n)$ bits.
  \item \textup{\RLBWT{}-LZM} with the majority tree of \lemmaref{lem:rle-majority}
    and cached predictions has update time $O(\log z+\log\log n)$,
    prediction time $O(1)$, and space $O(z\log z\log n)$ words.
  \item If $\Sigma=\{0,\dots,\sigma-1\}$ is known in advance, \textup{\RLBWT{}-LZP} with
    the alphabet-indexed structure of \lemmaref{lem:rle-indexed} and cached
    predictions has update time $O(\sigma(1+\log z/\log\log n))$, prediction
    time $O(1)$, and space $O(\sigma z\log z\log n)$ words.
\end{enumerate}
In particular, if $z=\polylog(n)$, \textup{\RLBWT{}-LZM} uses $\polylog(n)$ space and
$O(\log\log n)$ time per round; if moreover $\sigma=O(1)$, the predictor in
(3) uses $\polylog(n)$ space and $O(1)$ time per round.
\end{corollary}

\begin{proof}
By \lemmaref{lem:runs}, $r\le r_{\max}(x)=O(z\log z\log n)$ throughout, so
$\log r=O(\log z+\log\log n)$ and, since $\log w=\Theta(\log\log n)$,
$\log r/\log w=O(1+\log z/\log\log n)$. Every active symbol heads a run of
$L^-$, so the list of active symbols fits in the space bounds.
(1) follows from \theoremref{thm:lzp} and \lemmaref{lem:rle-trees}, and (2)
from \theoremref{thm:lzm} and \lemmaref{lem:rle-majority}, where caching is
free since $\Tcand=O(\log r)$. For (3), \theoremref{thm:lzp} with
\lemmaref{lem:rle-indexed} gives the update time
$O(\sigma\Trank+\Tcount+\Tins)=O(\sigma(1+\log r/\log w))$, and the space $O(\sigma(r_{\max}(x)+1))=O(\sigma z\log z\log n)$
words.
\end{proof}

\section{Discussion and Future Work}
\label{sec:open}

Our \RLBWT{} construction reduces the cost of our predictors to that of an
incremental RLE structure. The implementations of
\sectionref{sec:rle} are incomparable (\tableref{tab:comparison}): the
balanced-tree structures take $O(\log r)$ time per operation, while insertion
into the alphabet-indexed structure takes $O(\sigma(1+\log r/\log w))$ time.
Thus, for $r\ge w$, the alphabet-indexed structure is faster when
$\sigma=o(\log w)$, and slower when $\sigma=\omega(\log w)$. By
\theoremref{thm:lzm}, a single structure with the following guarantees would
give an LZM predictor that is at least as fast as every predictor in
\tableref{tab:comparison}, in both update and prediction time.

\begin{openproblem}
Is there a deterministic augmented incremental RLE structure
(\definitionref{def:rle-aug}), for an alphabet that is not known in advance,
that supports \emph{rank}, $C_L$, \emph{insert} and \emph{candidate} in
$O(1+\log r/\log w)$ worst-case time each, using $O(r)$ words?
\end{openproblem}

By \theoremref{thm:lzm} and \lemmaref{lem:runs}, such a structure would give
cached LZM with worst-case update time $O(1+\log z/\log\log n)$ and
prediction time $O(1)$ for $w=\Theta(\log n)$, with the same mistake and
space bounds as in \corollaryref{cor:main}. In particular, it would run
in constant time per round whenever $z=\polylog(n)$, with no assumption on
the alphabet.

The run lengths can already be maintained in $O(1+\log r/\log w)$ time per
operation (\lemmaref{lem:dps}), which is optimal for partial sums with small updates
\citep{patrascuLogarithmicLowerBounds2005}; the difficulty lies in the \emph{rank} queries on the run heads and in the \emph{candidate} queries.
\citet{munroCompressedDataStructures2015} support \emph{rank} queries and insertions on
a sequence of length $r$ over an alphabet fixed in advance in
$O(\log r/\log\log r)$ worst-case time. This is slower than required when $w$
is large compared with $r$: if $w=\Theta(\log n)$ and $z=\polylog(n)$, then
$r=\polylog(n)$ by \lemmaref{lem:runs}, and their bound does not give
constant time per round.

We believe that the answer is positive nevertheless. Storing the run heads
in a B-tree of fan-out $w^{\Theta(1)}$, as in the dynamic integer sets of
\citet{patrascuDynamicIntegerSets2014} and the partial sums of
\citet{billeDynamicRelativeCompression2018}, gives height
$O(1+\log r/\log w)$, so it would suffice to do $O(1)$ work per level.
The main obstacles are keeping the per-node information for these two
queries within linear space, and maintaining it with worst-case rather than
amortized bounds. We leave a construction for future work.

Finally, \RLBWT{}-LZM uses $O(z\log z\log^2 n)$ bits of space, while
Kosoy's implementation of LZP keeps only $O(z\log n)$ bits of persistent
state (\tableref{tab:comparison}); it would be interesting to close this gap
without slowing down the predictor.

\acks{I would like to thank Vanessa Kosoy for mentoring me throughout the project and for providing much helpful guidance.
This work was carried out during the Principles of Intelligent Behavior in Biological and Social Systems (PIBBSS) 2026 Summer Fellowship, and I thank PIBBSS for continued support of the research work.
It was also supported by the Advanced Research + Invention Agency (ARIA) of the United Kingdom and by Coefficient Giving in San Francisco, California.}

\section*{AI Methodology}
This work was developed through an extensive collaboration with AI systems.
The writing was drafted with substantial assistance from AI models (OpenAI ChatGPT and Anthropic Claude), based on my research notes and guidance, and I then further reviewed and corrected it.
The AI models also helped with proofs and literature review, and identified some errors in earlier drafts. I take full responsibility for the content of the paper.

\printbibliography

\appendix
\counterwithin*{theorem}{section}
\renewcommand{\thetheorem}{\thesection.\arabic{theorem}}
\renewcommand{\theHtheorem}{\thesection.\arabic{theorem}}
\counterwithin*{algocf}{section}
\renewcommand{\thealgocf}{\thesection.\arabic{algocf}}
\def\theHalgocf{\thesection.\arabic{algocf}}

\section{Proofs for Section~\ref{sec:setting}}
\label{app:setting}

\begin{proofof}{\lemmaref{lem:extension}}
If $H$ is empty, then so is $P$, $\occ_H(Pc)=0$, and the factorization of
$Hc=c$ is the single new phrase $c$, as claimed. Assume now that $H$ is
nonempty, and let $\ell\coloneqq|H|-|P|$ be the starting position of $P$. The phrases of
$H$ before $P$ are also phrases of $Hc$. Indeed, let $f_j$ be such a phrase,
starting at $\ell_j$. Then $f_j$ followed by the next symbol of $H$ has no
occurrence starting before $\ell_j$ in $H$, and hence none in $Hc$, since
such an occurrence would end before $\ell_j+|f_j|+1\le|H|$. So $f_j$ is
still the longest prefix with an earlier occurrence (or a new symbol). It
remains to find the phrases of $Hc$ starting at $\ell$.

An occurrence of $Pc$ starting at a position $j<\ell$ ends at position
$j+|P|\le\ell-1+|P|=|H|-1$, so it lies inside $H$. Conversely, every
occurrence of $Pc$ in $H$ starts at a position $j\le|H|-|P|-1<\ell$. Hence
$Pc$ has an occurrence starting before $\ell$ if and only if
$\occ_H(Pc)>0$. If it does, then $Pc=(Hc)[\ell:]$ is the whole remaining
suffix, and it is the phrase of $Hc$ starting at $\ell$ (note that if $P$ is
a new symbol, it has no earlier occurrence and $\occ_H(Pc)=0$). Otherwise,
the phrase starting at $\ell$ is still $P$, and the next phrase starts at
$|H|$. It consists of the single remaining symbol $c$, either as a new
symbol or as the longest prefix of $c$ with an earlier occurrence.
\end{proofof}

\begin{corollary}[Prefixes]
\label{cor:prefix}
Let $u\in\Sigma^*$ and $0\le k\le|u|$. The factorization of $u[:k]$ consists
of the phrases of $u$ that end by position $k$, followed by a prefix of the
next phrase (if $k$ falls inside it). In particular, $z(u[:k])\le z(u)$.
\end{corollary}

\begin{proof}
We show that the phrases of $u[:k]$ start exactly at the starting positions
of phrases of $u$ that are smaller than $k$. Fix $k$, and proceed by
induction on $|u|\ge k$. For $|u|=k$ there is nothing to prove. If $|u|>k$,
write $u=Hc$ with $c\in\Sigma$, so that $|H|\ge k$ and $H[:k]=u[:k]$. By
\lemmaref{lem:extension}, the phrases of $u$ start at the same positions as
the phrases of $H$, and possibly also at $|H|\ge k$. Hence $u$ and $H$ have
the same phrase starting positions below $k$, and by the induction
hypothesis for $H$, these are the starting positions of the phrases of
$u[:k]$.
\end{proof}

\begin{proofof}{\theoremref{thm:mistakes}}
Let $x=f_0\cdots f_{z-1}$ be the LZ77 factorization of $x$, and let $\Pi$ be
majority-respecting. We show that $\Pi$ makes at most $\log_2 n+1$ mistakes
while predicting the symbols of a single phrase $f=f_i$, which starts at
position $\ell\coloneqq|f_0\cdots f_{i-1}|$. We charge one mistake to its
first symbol $f[0]=x[\ell]$ (this also covers the prediction on the empty
history, for $i=0$). If $|f|=1$ we are done. Otherwise, $f$ has a source: a
position $p<\ell$ with $x[p:p+|f|]=f$. For $1\le k\le|f|$, let
\[
  V_k\coloneqq\bigl\{j<\ell\;\big|\;x[j:j+k]=f[:k]\bigr\},
\]
so that $p\in V_k$ for all $k$, and $|V_1|\le\ell<n$.

Fix $1\le k<|f|$, and consider the prediction of $x[\ell+k]=f[k]$, which is
made after the history $H\coloneqq x[:\ell+k]$. By
\corollaryref{cor:prefix}, the factorization of $H$ is
$f_0\cdots f_{i-1}$ followed by $f[:k]$, so the last phrase of $H$ is
$f[:k]$. An occurrence of $f[:k]\,a$ in $H$, for $a\in\Sigma$, has length
$k+1$ and therefore starts at a position $j\le\ell-1$. Conversely, for every
$j\in V_k$, the symbol $x[j+k]$ lies in $H$ since $j+k<\ell+k$. Hence
\[
  v_H(f[:k],a)=\bigl|\{j\in V_k\mid x[j+k]=a\}\bigr|
  \quad\text{for every }a\in\Sigma,
\]
so that the votes partition $V_k$ and $v_H(f[:k])=|V_k|$. Moreover,
$V_{k+1}=\{j\in V_k\mid x[j+k]=f[k]\}$, so $|V_{k+1}|=v_H(f[:k],f[k])$.

If $\Pi$ makes a mistake on $f[k]$, then $f[k]$ does not hold a strict
majority of the votes, since otherwise $\Pi$ would predict it. Therefore
$|V_{k+1}|\le|V_k|/2$. If $m$ mistakes are made on $f[1],\dots,f[|f|-1]$,
then
\[
  1\le|V_{|f|}|\le 2^{-m}|V_1|<2^{-m}n,
\]
so $m<\log_2 n$. Together with the first symbol, at most $\log_2 n+1$
mistakes are made on $f$, and summing over the $z$ phrases proves the
theorem.
\end{proofof}

\section{Proofs for Section~\ref{sec:rle}}
\label{app:rle}

Throughout, $L=c_1^{e_1}\cdots c_r^{e_r}$, and run $j$ occupies the positions
$[E_{j-1},E_j)$ of $L$, where $E_j\coloneqq e_1+\dots+e_j$.
\bigskip

\begin{proofof}{\lemmaref{lem:run-edits}}
Consider $\insertop(i,c)$. If $L[i-1]=c$ or $L[i]=c$ (for the positions that
exist), the new symbol extends the run containing that position. Otherwise,
if $L[i-1]$ and $L[i]$ belong to the same run, of a symbol $c'\ne c$, the
insertion splits it into two runs of $c'$ with the new run $c^1$ between
them. Otherwise, $i$ is a boundary between two runs of symbols other than
$c$, or an end of $L$, and the insertion creates a new run $c^1$ there. In
all cases, no two runs of the same symbol become adjacent, so no runs merge,
and no run disappears.
\end{proofof}

\begin{proofof}{\lemmaref{lem:rle-trees}}
We use the structure described in \sectionref{sec:rle-trees}: the
partial-sums structures $A$ and $A_c$ over the run lengths,
the run-head string $L'$, and the dictionary $F$ of active symbols.

\textbf{Components.} The structures $A$, $A_c$ and $F$ are
worst-case balanced search trees (e.g.\ red-black trees) with one node per
entry, augmented with subtree sizes and subtree sums. They support prefix
sums, searches, updates and insertions of entries in $O(\log r)$
worst-case time. The nodes of $A$ also store the run heads,
and the node of each active symbol $c$ in $F$ stores its frequency and a
pointer to $A_c$. Each node takes $O(1)$ words.

For the run-head string $L'$ we use
\citet[Theorem~10.5]{navarroOptimalDynamicSequence2014}: a dynamic string of
length $m$ over a general alphabet, with $s$ distinct symbols, can be stored
in $mH_0+o(m\log s)+O(s\log m)$ bits plus the space of a dictionary of its
symbols, supporting \emph{rank} in $O(T(s)+\log m)$ worst-case time and
insertions in $O(U(s)+\log m)$ worst-case time, where $T(s)$ and $U(s)$ are
the search and insertion times of the dictionary. We use a balanced search
tree as the dictionary, so that $T(s),U(s)=O(\log s)$ and new symbols can
be added at any time. Here $m=r$ and $s\le r$, so both operations take
$O(\log r)$ time, and the space is $O(r\log r)$ bits plus $O(r)$
words. The structure is described for words of $\Theta(\log m)$ bits. Since
$r\le n$ and $w=\Omega(\log n)$, each of its words fits in $O(1)$ of our
words, so it occupies $O(r)$ words, and each of its word operations can be
simulated in constant time. Its length-recalibration mechanism keeps the bounds worst-case as $r$
grows, including the construction of its lookup tables
\citep[Section~9.3]{navarroOptimalDynamicSequence2014}, and instances of
constant size can be handled directly.

\textbf{Queries.} $\rank_c(L,0)=0$, and if $c$ is not in $F$ then
$\rank_c(L,i)=0$ for all $i$. Otherwise, \equationref{eq:rle-rank} uses one
search and one prefix sum in $A$, one \emph{rank} query on
$L'$ and one prefix sum in $A_c$; the run head $L'[k]$ is stored in the node
of $A$ found by the search. The query $C_L(c)$ is a prefix sum over
the keys of $F$ smaller than $c$, which also works when $c$ is not active.
Both take $O(\log r)$ time.

\textbf{Insertions.} For $\insertop(i,c)$, we find the runs containing the
positions $i-1$ and $i$ (if they exist) by searches in $A$,
and their symbols in the nodes of $A$. If $c$ is not active,
we first create an empty tree $A_c$ and insert $c$ into $F$ (with frequency
$0$) and into the dictionary of $L'$.
We then perform the edit of \lemmaref{lem:run-edits}. The index of a run $j$ of $c$ among the runs of $c$ is $\rank_c(L',j-1)+1$, and similarly for other symbols.
\begin{itemize}
  \item Extending the run $j$ of $c$ next to position $i$ adds one to entry
    $j$ of $A$ and to entry $\rank_c(L',j-1)+1$ of $A_c$.
  \item Splitting run $j$, of a symbol $c'\ne c$, at offset
    $\tau\coloneqq i-E_{j-1}$ replaces entry $j$ of $A$ by the
    three entries $\tau,1,e_j-\tau$, and entry $\rank_{c'}(L',j-1)+1$ of
    $A_{c'}$ by the two entries $\tau,e_j-\tau$. It inserts a new entry
    $\rank_c(L',j)+1$, equal to $1$, into $A_c$, and inserts the two symbols
    $c\,c'$ after the first $j$ symbols of $L'$.
  \item A new run $c^1$ after the first $j$ runs ($0\le j\le r$) is inserted
    as a new entry $j+1$ of $A$ and a new entry
    $\rank_c(L',j)+1$ of $A_c$, both equal to $1$, and $c$ is inserted after
    the first $j$ symbols of $L'$ (at the beginning if $j=0$).
\end{itemize}
Finally, the frequency of $c$ in $F$ grows by one. These are $O(1)$
operations of $O(\log r)$ time each.

\textbf{Space.} If $c$ has $r_c$ runs, then $A_c$ has $r_c$ nodes and its
root is referenced from the node of $c$ in $F$; since $\sum_c r_c=r$, the
trees $A$, $A_c$ and $F$ occupy $O(r)$ words, as does $L'$.
Their operations use $O(\log r)$ words of temporary storage, which is also
$O(r)$.
\end{proofof}

\begin{proofof}{\lemmaref{lem:dps}}
\citet[Theorem~2]{billeDynamicRelativeCompression2018} prove a more general
result with a parameter $\delta$. Their structure maintains a sequence of $s$
$w$-bit integers in linear space, and supports $\sumop$ and $\searchop$,
changing an entry by less than $2^{\delta}$, inserting and deleting entries
smaller than $2^{\delta}$, merging two consecutive entries, and dividing an
entry, in $O(1+\log s/\log(w/\delta))$ time per operation; the proof in their
Section~3.2 gives worst-case bounds. For $\delta=1$, adding one to an entry
and inserting an entry equal to $0$ or $1$ are among these operations, and
the time per operation is $O(1+\log s/\log w)$.
\end{proofof}

\begin{proofof}{\lemmaref{lem:rle-indexed}}
We use the structure described in \sectionref{sec:rle-indexed}. It
consists of the structures $A$ and $W_c$ of \lemmaref{lem:dps}, where $A$
stores the run lengths and $W_c$, for each of the $\sigma$ symbols $c$,
stores the lengths of the runs of $c$ and zeros for the other runs, and of
the array $\mathrm{cnt}$ with $\mathrm{cnt}[y]=C_L(y)$ for every symbol $y$. Initially, these are $\sigma+1$ empty
structures and $\sigma$ zeros. The entries of $A$ are positive, so
$A.\searchop(i)$ is the run containing $L[i-1]$ for $0<i\le N$, and the
formula for $\rank_c(L,i)$ in \sectionref{sec:rle-indexed} uses four
operations of \lemmaref{lem:dps} on sequences of length $r$. This takes
$O(1+\log r/\log w)$ time. The value $C_L(c)=\mathrm{cnt}[c]$ takes $O(1)$
time.

For $\insertop(i,c)$, we find the runs containing the positions $i-1$ and
$i$ (if they exist) by searches in $A$, and the symbol of each such run $j$
by testing $W_y.\sumop(j)>W_y.\sumop(j-1)$ for all $y\in\Sigma$. This takes
$O(\sigma)$ operations of \lemmaref{lem:dps}. We then perform the edit of \lemmaref{lem:run-edits}
in all $\sigma+1$ structures at once:
\begin{itemize}
  \item extending the run $j$ of $c$ next to position $i$ adds one to entry
    $j$ of $A$ and of $W_c$;
  \item splitting run $j$, of a symbol $c'\ne c$, at offset
    $\tau\coloneqq i-E_{j-1}$ divides entry $j$ into $\tau$ and $e_j-\tau$
    in $A$ and in $W_{c'}$, and into two zeros in every other $W_y$; it then
    inserts a new entry $j+1$, equal to $1$ in $A$ and in $W_c$ and to $0$
    in every other $W_y$;
  \item a new run $c^1$ after the first $j$ runs is inserted as a new entry
    $j+1$, equal to $1$ in $A$ and in $W_c$ and to $0$ in every other $W_y$.
\end{itemize}
Finally, $\mathrm{cnt}[y]$ grows by one for every $y>c$, in $O(\sigma)$
time. The edit uses $O(1)$ operations of \lemmaref{lem:dps} in each of the
$\sigma+1$ structures, so an insertion takes $O(\sigma)$ such operations
and $O(\sigma(1+\log r/\log w))$ time. By \lemmaref{lem:dps}, each of the $\sigma+1$ sequences takes
$O(r+1)$ words. Together with the $\sigma$-entry array $\mathrm{cnt}$,
the space is $O(\sigma(r+1))$ words.
\end{proofof}

\begin{proofof}{\lemmaref{lem:rle-majority}}
We augment every node $v$ of the balanced tree $A$ of
\lemmaref{lem:rle-trees} with the summary
\[
  \operatorname{summary}(v)\coloneqq\bigl(\operatorname{summary}(v_{\mathrm L})\star(c_v,e_v)\bigr)
  \star\operatorname{summary}(v_{\mathrm R}),
\]
where $c_v^{e_v}$ is the run stored at $v$, $v_{\mathrm L}$ and
$v_{\mathrm R}$ are its children, and the summary of an empty subtree is
$\bot$. By \lemmaref{lem:summaries} and induction, $\operatorname{summary}(v)$ is valid
for the multiset of symbols in the runs of the subtree of $v$. A summary takes two words, since its balance is at most $N<2^w$.
It depends only on the fields of $v$ and of its children. Hence, as for subtree sums, it can be maintained under insertions, including rebalancing, in $O(\log r)$ worst-case time.

\algorithmref{alg:candidate} answers a query $\cand(a,b)$. It handles the
two partial runs at the ends of $L[a:b]$ separately, so that its recursive
procedures only deal with whole runs. These procedures use the subtree sizes
stored in $A$: $\operatorname{size}(v)$ is the number of runs in the subtree
of $v$ (zero for an empty subtree), and the \emph{rank} of a run in this
subtree is the number of runs before it there, so that the ranks are
$0,\dots,\operatorname{size}(v)-1$ in the order of $L$. As $\star$ is not
associative, we read $s_1\star s_2\star s_3$ as $(s_1\star s_2)\star s_3$;
any other order would also do.

\begin{algorithm2e}[htb]
\caption{The query $\cand(a,b)$ in the augmented tree $A$}
\label{alg:candidate}
\SetKwFunction{FnRuns}{Runs}
\SetKwFunction{FnPrefix}{Prefix}
\SetKwFunction{FnSuffix}{Suffix}
\Proc{$\cand(a,b)$}{
  \lIf{$a=b$}{\Return $\bot$}
  find the runs $j_a$ and $j_b$ containing $L[a]$ and $L[b-1]$, by searches in $A$\;
  \lIf{$j_a=j_b$}{\Return $c_{j_a}$}
  $s\gets(c_{j_a},E_{j_a}-a)\star\FnRuns{$\mathit{root},j_a,j_b-1$}\star(c_{j_b},b-E_{j_b-1})$\;
  \lIf{$s=\bot$}{\Return $\bot$}
  \lElse{\Return the symbol $c$ of $s=(c,p)$}
}
\BlankLine
\Proc{\FnRuns{$v,x,y$}\tcp*[f]{the runs of ranks $x,\dots,y-1$ in the subtree of $v$}}{
  \lIf{$x=y$}{\Return $\bot$}
  $k\gets\operatorname{size}(v_{\mathrm L})$\tcp*{the rank of the run of $v$}
  \lIf{$y\le k$}{\Return $\FnRuns{$v_{\mathrm L},x,y$}$}
  \lIf{$x>k$}{\Return $\FnRuns{$v_{\mathrm R},x-k-1,y-k-1$}$}
  \Return $\FnSuffix{$v_{\mathrm L},x$}\star(c_v,e_v)\star\FnPrefix{$v_{\mathrm R},y-k-1$}$\tcp*{$x\le k<y$}
}
\BlankLine
\Proc{\FnPrefix{$v,y$}\tcp*[f]{the runs of ranks $0,\dots,y-1$}}{
  \lIf{$y=0$}{\Return $\bot$}
  $k\gets\operatorname{size}(v_{\mathrm L})$\;
  \lIf{$y\le k$}{\Return $\FnPrefix{$v_{\mathrm L},y$}$}
  \Return $\operatorname{summary}(v_{\mathrm L})\star(c_v,e_v)\star\FnPrefix{$v_{\mathrm R},y-k-1$}$\;
}
\BlankLine
\Proc{\FnSuffix{$v,x$}\tcp*[f]{the runs of ranks $x,\dots,\operatorname{size}(v)-1$}}{
  \lIf{$x=\operatorname{size}(v)$}{\Return $\bot$}
  $k\gets\operatorname{size}(v_{\mathrm L})$\;
  \lIf{$x>k$}{\Return $\FnSuffix{$v_{\mathrm R},x-k-1$}$}
  \Return $\FnSuffix{$v_{\mathrm L},x$}\star(c_v,e_v)\star\operatorname{summary}(v_{\mathrm R})$\;
}
\end{algorithm2e}

\emph{Correctness.} Every call has $0\le x\le y\le\operatorname{size}(v)$
for \textsc{Runs}, $0\le y\le\operatorname{size}(v)$ for \textsc{Prefix}, and
$0\le x\le\operatorname{size}(v)$ for \textsc{Suffix}, so a call on an empty
subtree returns $\bot$ in its first line. In the other cases, the runs in
question either lie in one subtree of $v$, where the ranks in the right
subtree are shifted by $k+1$, or consist of a suffix of the runs of
$v_{\mathrm L}$, the run of $v$, and a prefix of the runs of $v_{\mathrm R}$.
By induction on the height of $v$ and \lemmaref{lem:summaries}, each
procedure returns a valid summary of the multiset of symbols in its runs. In
$\cand(a,b)$ with $j_a<j_b$, the string $L[a:b]$ consists of the last
$E_{j_a}-a$ symbols of run $j_a$, the runs $j_a+1,\dots,j_b-1$, which have
ranks $j_a,\dots,j_b-2$ in $A$, and the first $b-E_{j_b-1}$ symbols of run
$j_b$. Hence $s$ is valid for $L[a:b]$, and by \lemmaref{lem:summaries} it is
$(c,p)$ with $p\ge1$ whenever $c$ has a strict majority in $L[a:b]$.

\emph{Time.} Every call of \textsc{Prefix} or \textsc{Suffix} takes $O(1)$
time besides its recursive call, of which there is at most one, on a child of
its node. Hence each of them visits the nodes of one downward path in $A$.
The same holds for \textsc{Runs}, except in its last case, at a node $u$,
where it calls \textsc{Suffix} on $u_{\mathrm L}$ and \textsc{Prefix} on
$u_{\mathrm R}$ once each and stops. So the nodes visited by a query lie on
two downward paths from the root, which agree down to $u$ and each have
$O(\log r)$ nodes, besides the two searches. It takes $O(\log r)$ time and, for the recursion,
$O(\log r)$ words of temporary space. The augmentation adds $O(1)$ words per
node.
\end{proofof}

\section{Proofs for Section~\ref{sec:predictors}}
\label{app:predictors}

\begin{proofof}{\lemmaref{lem:sentinel}}
(1) The string $L^-$ is $L$ without the cell $p_\$$: the positions before
$p_\$$ keep their index, and the index of each position after it decreases
by one. This
gives the formula for $\rank_c$ with $c\in\Sigma$. Since $\$<c$, $C_L(c)$
counts the sentinel in addition to the symbols of $L^-$ that are smaller
than $c$.

(2) The suffixes of $cY$ are $cY$ itself and the suffixes of $Y$, whose
relative order does not change. The \BWT{} symbol of the row of $Y$, which
is the row $p_\$$, changes from $\$$ to $c$; in $L^-$, this inserts $c$ at
position $p_\$$, since the cells of $L$ before $p_\$$ are the first $p_\$$
cells of $L^-$. The new suffix $cY$ has \BWT{} symbol $\$$, so its row
changes $p_\$$ but not $L^-$. Its row is the number of suffixes of $Y$ that
are smaller than $cY$. These are the suffixes starting with a symbol smaller
than $c$, of which there are $C_L(c)$ (the first symbols of the suffixes of
$Y$ are exactly the symbols of $L$), and the suffixes $cZ$ with $Z<Y$. The
latter correspond to the suffixes $Z$ of $Y$ whose row precedes $p_\$$ and
whose \BWT{} symbol is $c$, of which there are $\rank_c(L,p_\$)$. By (1),
$C_L(c)+\rank_c(L,p_\$)=s$. The costs follow directly.
\end{proofof}

\begin{proofof}{\lemmaref{lem:votes}}
If $H$ is empty, then $P=\varepsilon$, $Y=L=\$$, $I=[0,1)$ and $p_\$=0$.
All $\nu_c$ and $v_H(P,c)$ are zero, and $b-a-1=0$, so the claims hold.
Assume now that $H$ is nonempty, and write $t=|H|$ and $k=|P|\ge1$.
The occurrences $H[j:j+k]=P$ correspond
bijectively to the occurrences of $\rev{P}$ in $Y=\rev{H}\$$, at the
positions $q=t-j-k$. If $j+k<t$, then $q>0$, and the symbol following the
occurrence in $H$ is $H[j+k]=Y[q-1]$, which is the \BWT{} symbol of the row
of the suffix $Y[q:]$. So, for $c\in\Sigma$, the rows of $[a,b)$ with \BWT{}
symbol $c$ correspond to the occurrences of $P$ in $H$ that are followed by
$c$, and $\nu_c=\occ_H(Pc)=v_H(P,c)$. The remaining occurrence, with $j+k=t$,
is the suffix $P$ of $H$. It has $q=0$, so its row is the row of $Y$ itself,
namely $p_\$$, whose \BWT{} symbol is $\$$. Hence $a\le p_\$<b$, and the
other $b-a-1$ rows of the interval have symbols in $\Sigma$.
\end{proofof}

\begin{proofof}{\theoremref{thm:lzp}}
We show by induction that after every update, the structure stores $L^-$
and $p_\$$ for $Y=\rev{H}\$$, the list contains the active symbols of $H$,
and $I=I_Y(\rev{P})$ for the last phrase $P$ of $H$. The first two claims
follow from \lemmaref{lem:sentinel}, and from the fact that $c$ is new if and
only if $\rank_c(L,N)=0$, since $L$ and $Y$ contain the same symbols.

Initially, $I=[0,1)=I_\$(\varepsilon)$. On an update, $J$ is nonempty if
and only if $c\rev{P}$ occurs in $Y$. Since
$c\rev{P}$ does not contain $\$$, this happens if and only if $Pc$ occurs in
$H$. By \lemmaref{lem:extension}, the new last phrase is therefore $Pc$ if
$J$ is nonempty, and $c$ otherwise. In the first case, the interval of
$c\rev{P}$ in the new text consists of the rows of the old suffixes in $J$
and of the new suffix $cY$, which starts with $c\rev{P}$. The rows of the old
suffixes keep their order, and the index of each one after the new row $s$
increases by one; hence the interval is $[a',b'+1)$. In the second case, the
interval $I_{cY}(c)$ is computed directly.

By \lemmaref{lem:votes}, the values $\nu_c$ are the votes of LZP. On a
nonempty history the list of active symbols is nonempty, and since only
active symbols can have a positive vote, an active symbol with the largest
$\nu_c$ maximizes $v_H(P,c)$ over $\Sigma$, as LZP requires. On the empty
history all symbols are tied. The
mistake bound then follows from \theoremref{thm:mistakes}, since LZP is
majority-respecting.

An update performs at most two backward steps (each one $C_L$ query and two
\emph{rank} queries), one \emph{rank} query for the test of a new symbol, and
one prepend, which takes $O(\Trank+\Tcount+\Tins)$ time by
\lemmaref{lem:sentinel}. A prediction computes two \emph{rank} queries for
each active symbol. With caching, the update also computes the next
prediction. Besides the structure, the state consists of $O(1)$ words and
the list of active symbols.
\end{proofof}

\begin{proofof}{\theoremref{thm:lzm}}
The update and the invariant are those of \RLBWT{}-LZP. Deleting the row
$p_\$\in[a,b)$ from $L$ leaves the other rows of
$[a,b)$ at the positions $[a,b-1)$ of $L^-$. By \lemmaref{lem:votes}, the
multiset of symbols of $L^-[a:b-1]$ contains each $c\in\Sigma$ exactly
$v_H(P,c)$ times, and has size $b-a-1=v_H(P)$. If
$v_H(P,c)>\tfrac{1}{2}v_H(P)$, then $c$ occurs in $L^-[a:b-1]$ more than
$(b-a-1)/2$ times, so the candidate query returns $c$, which is predicted.
Hence the predictor is majority-respecting, and \theoremref{thm:mistakes}
applies. A prediction makes one candidate query, and the state has $O(1)$
words besides the structure.
\end{proofof}

\begin{proofof}{\lemmaref{lem:runs}}
For a string $u$ and $k\ge1$, let $d_k(u)$ be the number of distinct
substrings of $u$ of length $k$. The substring complexity of $u$ is
$\delta(u)\coloneqq\max_{1\le k\le|u|}d_k(u)/k$. We prove the stronger bound
\[
  r_{\max}(x)=O(\delta(x)\log\delta(x)\log n),
\]
which implies the lemma since $\delta(x)\le z(x)$
(see \citealp[Sections~3.3, 3.9 and~3.10]{navarroIndexingHighlyRepetitive2022}).

Let $T$ be a text of length $m$ whose last symbol is a unique smallest
sentinel $\$$. \citet{kempaResolutionBurrowsWheelerTransform2020} use a
cyclic version of the substring complexity: for $k\ge1$, let $d'_k(T)$ be the
number of distinct substrings of length $k$ of the infinite string
$TTT\cdots$, and let $\delta'(T)\coloneqq\max_{1\le k\le m}d'_k(T)/k$. For
$k\le m$, every substring of $T$ of length $k$ is also a substring of
$TTT\cdots$, and a substring of $TTT\cdots$ of length $k$ that is not a
substring of $T$ contains $\$$ at one of its first $k-1$ positions, and is
determined by that position. Hence $d_k(T)\le d'_k(T)\le d_k(T)+k-1$, and
$\delta(T)\le\delta'(T)\le\delta(T)+1$. Writing $\delta'$ for $\delta'(T)$,
\citet[Theorem~III.7]{kempaResolutionBurrowsWheelerTransform2020} show that
\[
  r(\BWT(T))=O\Bigl(\delta'\log\delta'\cdot
  \max\Bigl\{1,\log\frac{m}{\delta'\log\delta'}\Bigr\}\Bigr)
  =O(\delta'\log\delta'\log m),
\]
and their \BWT{} coincides with ours.

Let $T\coloneqq\rev{x[:t]}\$$ for some $1\le t\le n$ (for $t=0$, the \BWT{}
has one run). Then $\delta(T)\ge2$, since $T$ has at least two distinct
symbols. The only substring of $T$ of length $k$ that contains $\$$ is its
suffix, so $d_k(T)\le d_k(\rev{x[:t]})+1$ and
$\delta(T)\le\delta(\rev{x[:t]})+1$. Moreover, $\delta$ is invariant under
reversal, and $\delta(x[:t])\le\delta(x)$ since every substring of $x[:t]$ is
a substring of $x$. Hence $2\le\delta'\le\delta(x)+2\le3\delta(x)$, and
$r(\BWT(T))=O(\delta(x)\log\delta(x)\log n)$, using $m=t+1\le n+1$.
\end{proofof}

Consequently, the time and space bounds in \corollaryref{cor:main} also hold
with $z(x)$ replaced by $\delta(x)\le z(x)$; the mistake bound remains
$z(x)(\log_2 n+1)$.

\end{document}